\documentclass{ws-procs975x65}
\usepackage{graphicx,amsmath}
\begin{document}
\title{Enhanced Higgs diphoton rate and isospin symmetric Higgs boson}
\author{Michio Hashimoto$^*$}
\address{
   Chubu University, \\
   1200 Matsumoto-cho, Kasugai-shi, \\
   Aichi, 487-8501, JAPAN \\
   $^*$E-mail: michioh@isc.chubu.ac.jp}
\begin{abstract}
Based on Ref.~1, we introduce a model with an isospin symmetric Higgs boson
and study the properties of this particle, including the enhancement of 
its diphoton decay rate.
The predictions of the model relevant for future experiments are 
also discussed.
\end{abstract}
\date{\today}
\bodymatter
\section{Introduction}

Recently, the ATLAS and the CMS Collaborations discovered 
a new boson $h$ in the mass range 125--126~GeV.\cite{LHC}
The observed diphoton decay rate is about 1.6 times larger than
that of the Standard model (SM) Higgs boson $H$,
while the other properties seem rather similar to the SM.
This deviation in the diphoton channel, if established,
would be an indication of a new physics beyond the SM.

Based on Ref.~\refcite{Hashimoto:2012qe},
we show that the ATLAS and CMS data for 
the enhanced Higgs diphoton branching ratio can be explained 
in the class of models with isospin symmetric (IS) electroweak Higgs boson.
We also discuss the predictions of these models, 
which can be checked at the LHC in the near future.

\section{IS Higgs Model and Higgs decay rates}

The main characteristics of the IS Higgs boson models
are the following.\cite{Hashimoto:2012qe,Hashimoto:2009xi}
a) It is assumed that the dynamics primarily responsible for 
the electroweak symmetry breaking (EWSB) leads to the mass 
spectrum of quarks with no (or weak) isospin violation.
{\it Moreover, it is assumed that the values of these masses are 
of the order of the observed masses of the down-type quarks.} 
b) The second (central) assumption is introducing 
the horizontal interactions for the quarks in the three families. 
As a first step, a {\it subcritical} (although nearcritical, i.e., strong) 
diagonal horizontal interactions for the top quark is utilized 
which lead to the observed ratio $\frac{m_t}{m_b} \simeq 41.5$.\cite{pdg}
The second step is introducing {\it equal} strength horizontal 
flavor-changing-neutral (FCN) interactions between 
the $t$ and $c$ quarks and the $b$ and $s$ ones. 
As was shown in Ref.~\refcite {Hashimoto:2009xi}, 
these interactions naturally provide the observed ratio 
$m_c/m_s \simeq 13.4$ in the second family.\cite{pdg} 
As to the mild isospin violation in the first family, 
it was studied together with the effects of the family mixing, 
reflected in the Cabibbo-Kobayashi-Maskawa (CKM) matrix.\cite{Hashimoto:2009xi}

In this scenario, 
the main source of the isospin violation is only 
the strong top quark interactions. 
However, because these interactions are subcritical, 
the top quark plays a minor role in EWSB.
This distinguishes the IS Higgs scenario from the top quark condensate 
model.\cite{Miransky:1988xi,Nambu,Marciano:1989xd,Bardeen:1989ds,Hashimoto:2000uk}

One of the signatures of this scenario is the appearance of a
composite top-Higgs boson $h_t$
composed of the quarks and
antiquarks of the third family.\cite{Hashimoto:2009xi}
Note that unlike the topcolor assisted technicolor model 
(TC2)\cite{Hill:1994hp},
this class of models utilizes subcritical dynamics
for the top quark, so that the top-Higgs $h_t$ is heavy in general.
Here we also emphasize that while the top-Higgs boson $h_t$ has 
a large top-Yukawa coupling,
the IS Higgs boson $h$ does not, $y_t \simeq y_b \sim 10^{-2}$. 
On the other hand, the $hWW^{*}$ and $hZZ^{*}$ coupling constants are
close to those in the SM.
Also, the mixing between $h$ and much heavier $h_t$ should be small. 

Let us now describe the decay processes of the IS Higgs $h$.

It is well known that the $W$-loop contribution to $H \to \gamma\gamma$
is dominant in the SM, while the top-loop effect is destructive
against the $W$-loop. 
In the IS Higgs model, however, the Yukawa coupling between the top and 
the IS Higgs $h$ is as small as the bottom Yukawa coupling,
so that the top-loop contribution is strongly suppressed.
The partial decay width of $h \to \gamma\gamma$ is thus enhanced 
without changing essentially $h \to ZZ^{*}$ and $h \to WW^{*}$.
A rough estimate taking the isospin symmetric 
top and bottom Yukawa couplings $y_t \simeq y_b \approx 10^{-2}$
is as follows:
\begin{equation}
\label{gamma}  
  \frac{\Gamma^{\rm IS} (h \to \gamma \gamma)}
       {\Gamma^{\rm SM}(H \to \gamma \gamma)} \simeq 1.56 , \quad
  \frac{\Gamma^{\rm IS} (h \to WW^{*})}
       {\Gamma^{\rm SM}(H \to WW^{*})} = 
  \frac{\Gamma^{\rm IS} (h \to ZZ^{*})}
       {\Gamma^{\rm SM}(H \to ZZ^{*})} = 
       \left(\frac{v_h}{v}\right)^2 \simeq 0.96 . 
\end{equation}
Here using the Pagels-Stokar formula \cite{PS}, we estimated 
the vacuum expectation value (VEV) of the top-Higgs $h_t$ 
as $v_t = 50$ GeV, and
the VEV $v_h$ of the IS Higgs $h$ is given by the relation $v^2 = v_h^2 + v_t^2$
with $v = 246$ GeV. Note that
the values of the ratios in Eq. (\ref{gamma}) are not very sensitive to 
the value of $v_t$, e.g., for $v_t=40$--$100$ GeV,
the suppression factor in the pair decay modes to $WW^*$ and $ZZ^*$
is $0.97$--$0.84$ and 
the enhancement factor in the diphoton channel is $1.58$--$1.37$. 
For the decay mode of $h \to Z\gamma$, this model yields 
\begin{equation}
\label{Zgamma}
  \frac{\Gamma^{\rm IS} (h \to Z \gamma)}
       {\Gamma^{\rm SM}(H \to Z \gamma)} \simeq 1.07 \, .
\end{equation}

The values in Eq. (\ref{gamma}) agree well with the data in 
the ATLAS and CMS experiments.
However, obviously, the main production mechanism of the Higgs boson,
the gluon fusion process $gg \to h$, is now in trouble.
The presence of new chargeless colored particles, which considered by
several authors \cite{adj-S}
can help to resolve this problem.
We pursue this possibility in the next section.

\section{Benchmark Model with colored scalar}
\label{3}

As a benchmark model, 
we may introduce a real scalar field $S$ in the adjoint
representation of the color $SU(3)_c$:
\begin{eqnarray}
  {\cal L} \supset 
  {\cal L}_S = \frac{1}{2} (D_\mu S)^2 - \frac{1}{2} m_{0,S}^2 S^2 
  - \frac{\lambda_S}{4} S^4
  - \frac{\lambda_{h S}}{2} S^2 \Phi_h^\dagger \Phi_h,
  \label{Lag-S}
\end{eqnarray}
where $\Phi_h$ represents the IS Higgs doublet.
The effective Lagrangian ${\cal L}$ also contains
the IS Higgs quartic couplings $\lambda_h$, 
${\cal L} \supset - \lambda_h |\Phi_h|^4$.
The IS Higgs mass is $m_h = \sqrt{2\lambda_h} v_h$, and 
we will take it to be equal to $125$ GeV. 
The mass-squared term for the scalar $S$ is 
$  M_S^2 = m_{0,S}^2 + \frac{\lambda_{hS}}{2} v_h^2$,
and should be positive in order to avoid the color symmetry breaking.
Typically, $M_S \sim 200$~GeV. 

Taking into account the $S$ contribution to $gg \to h$,
we find appropriate values of the Higgs-portal coupling,
\begin{equation}
  \lambda_{hS} \simeq 2.5 \mbox{--} 2.7 \times \frac{M_S^2}{v v_h} \,.
\end{equation}
As a typical value, we may take $\lambda_{hS} = 1.8$ for 
$M_S= 200$~GeV and $v_t=50$~GeV.

A comment concerning the IS Higgs  quartic coupling $\lambda_h$ is in order.
In the SM, the Higgs mass 125~GeV suggests that the theory is 
perturbative up to an extremely high energy scale~\cite{RGE-lam}.
On the contrary, in the present model, when we take a large Higgs-portal 
coupling $\lambda_{hS}$ that reproduces $gg \to h$ correctly, 
the quartic coupling $\lambda_h$ will grow because
the $\beta$-function for $\lambda_h$ contains the $\lambda_{hS}^2$ term.
Also, there is no large negative contribution to 
the $\beta$-function for $\lambda_h$ from 
the top-Yukawa coupling $y_t \sim 10^{-2}$.

\begin{figure}[t]
  \begin{center}
  \includegraphics[width=7.5cm]{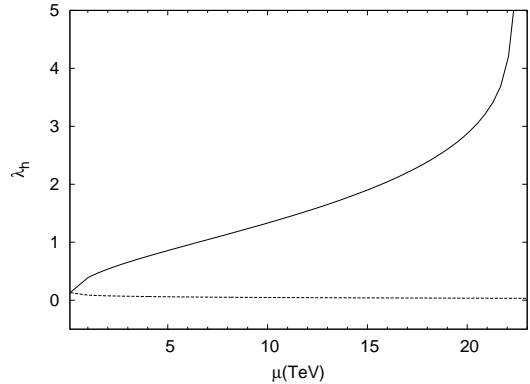}
   \end{center}
   \caption{The running behavior of the IS Higgs quartic coupling $\lambda_h$.
   The solid and dashed lines correspond to $\lambda_h$
   and the SM Higgs quartic coupling, respectively.
   We fixed the IS Higgs mass $m_h = \sqrt{2\lambda_h} v_h = 125$~GeV and
   took $\lambda_{hS}=1.8$ and $\lambda_S=1.5$.
   Unlike the SM, the IS Higgs quartic coupling grows up due to
   a large Higgs-portal coupling $\lambda_{hS}$  and a small
   top-Yukawa coupling $y_t$.
   \label{lam}}
\end{figure}

One can demonstrate such a behavior more explicitly by using the 
renormalization group equations.
In Fig.~\ref{lam}, the running of the coupling $\lambda_h$ is shown.
Taking a large Higgs-portal coupling $\lambda_{hS}=1.8$ and 
the $S^4$-coupling $\lambda_S=1.5$, 
it turns out that the coupling $\lambda_h$ rapidly grows.
The blowup scale strongly depends on the initial values of $\lambda_{hS}$ 
and $\lambda_S$.
A detailed analysis will be performed elsewhere.

Last but not least, we would like to mention that 
other realizations of the enhancement of the $h$ production 
are also possible.

\section{Conclusion}
\label{4}

We studied the properties of the IS Higgs boson.
The IS Higgs model can explain the enhanced Higgs diphoton decay rate 
observed at the LHC, 
and also makes several predictions. The most important of them is that
the value of the top-Yukawa coupling $h$-$t$-$\bar{t}$ should be close 
to the bottom-Yukawa one.
Another prediction relates to the decay mode $h \to Z\gamma$, 
which is enhanced only slightly, 
$\Gamma^{\rm IS} (h \to Z \gamma) = 1.07 \times \Gamma^{\rm SM} (H \to Z\gamma)$,
unlike $h \to \gamma\gamma$.
Last but not least, the LHC might potentially discover 
the top-Higgs resonance $h_t$, if lucky.
For details, see Ref.~\refcite{Hashimoto:2012qe}.

\end{document}